\documentclass{iaus}
\usepackage{graphicx}

  \checkfont{eurm10}
  \iffontfound
    \IfFileExists{upmath.sty}
      {\typeout{^^JFound AMS Euler Roman fonts on the system,
                   using the 'upmath' package.^^J}%
       \usepackage{upmath}}
      {\typeout{^^JFound AMS Euler Roman fonts on the system, but you
                   dont seem to have the}%
       \typeout{'upmath' package installed. iaus.cls can take advantage
                 of these fonts,^^Jif you use 'upmath' package.^^J}%
      }
  \else
  \fi

  \checkfont{msam10}
  \iffontfound
    \IfFileExists{amssymb.sty}
      {\typeout{^^JFound AMS Symbol fonts on the system, using the
                'amssymb' package.^^J}%
       \usepackage{amssymb}%

      }{}
  \fi

  \IfFileExists{amsbsy.sty}
    {\typeout{^^JFound the 'amsbsy' package on the system, using it.^^J}%
     \usepackage{amsbsy}}
    {\providecommand\boldsymbol[1]{\mbox{\boldmath $##1$}}}

\newsavebox{\astrutbox}
\sbox{\astrutbox}{\rule[-5pt]{0pt}{20pt}}

\title[The Interplay among Black Holes, Stars and ISM in Galactic 
       Nuclei]{High resolution X-ray spectra of quasars}

\author[S. Kaspi]%
{Shai Kaspi$^{1,2}$%
}

\affiliation{$^1$School of Physics \& Astronomy and the Wise Observatory,
The Raymond and Beverly Sackler Faculty of Exact Sciences,
Tel-Aviv University,
Tel-Aviv 69978, Israel email: shai@wise.tau.ac.il\\[\affilskip]
$^2$Physics Department, Technion, Haifa 32000, Israel}

\pubyear{2004}
\volume{222}
\pagerange{1--4}
\date{?? and in revised form ??}
\jname{The Interplay among Black Holes, Stars and ISM \\in Galactic Nuclei}
\editors{Th. Storchi Bergmann, L.C. Ho \& H.R. Schmitt, eds.}
\begin{document}

\maketitle

\begin{abstract}
Past X-ray observations by {\it ASCA} suggest that warm absorbers
(O\,{\sc vii} and O\,{\sc viii} edges) are apparently rare in
high luminosity AGNs (quasars) while they are more common in low
luminosity AGNs (Seyferts). However, this could be a selection
effect if high luminosity AGNs have mostly narrow absorption lines
(with no strong bound free edges), which escaped detection by
the low resolution of {\it ASCA}. To check this hypothesis we are
studying the high-resolution X-ray spectra of quasars from grating
spectrometers on board {\it Chandra} and {\it XMM-Newton} in search
for absorption lines. In this contribution we present spectra of
three quasars. The spectra show narrow (several hundred km\,s$^{-1}$)
absorption and emission X-ray lines from H-like and He-like ions of O,
Ne, Mg, and other abundant elements.  We also detect absorption from
iron L-shell lines and iron M-shell unresolved transition array. We
present the analysis of MR2251-178 where we find that at least two,
and probably three, distinct warm absorbers are needed to explain
the high resolution spectrum of this object.  We re-analyze the
high-resolution X-ray spectrum of PG\,1211+143 and suggest that an
outflow velocity of $\sim 3000$ km\,s$^{-1}$ provides an adequate
explanation to these data. We also present preliminary results form
the {\it Chandra}/HETGS observation of the quasar 4C74.26.
\end{abstract}

\firstsection 
\section{Introduction}

A significant fraction (50--75\%) of Seyfert-I Active Galactic Nuclei
(AGNs) exhibit absorption features due to highly ionized gas (so called
``warm absorber'' - WA) along the line-of-sight to the central source
(e.g., George et al. 1998). Early {\it ASCA} spectra showed only the
strongest bound-free O\,{\sc vii} and O\,{\sc viii} edges, while the
full complexity of the absorption spectrum are clearly evident in
recent {\it Chandra} and {\it XMM-Newton} observations. The X-ray
grating spectrometers aboard these two observatories reveal narrow
absorption lines from H-like and He-like ions as well as from lower
ionization ions of all abundant elements from carbon to iron, and
M-shell and L-shell absorption lines from iron ions. These narrow
absorption lines were found in most of Seyfert I AGNs observed so far
(e.g., NGC\,5548 - Steenbrugge et al. 2003; NGC\,4051 - Collinge et
al. 2001).  In most cases the absorption lines are blueshifted by
about several hundred km\,s$^{-1}$ relative to the systemic velocity,
indicating that the ionized absorbing gas is outflowing.

The best example, so far, is the Seyfert-I galaxy NGC\,3783, which has
the best observed X-ray spectrum in terms of spectral resolution and
S/N. It was observed for 900 ks with the High Energy Transmission
Grating Spectrometer (HETGS) on {\it Chandra} (Kaspi et al. 2002;
Netzer et al. 2003) and for about 280 ks with the Reflection Grating
Spectrometer (RGS) on {\it XMM-Newton} (Behar et al. 2003). The
spectrum shows more than 150 absorption lines from more than 30
ions spanning a wide range of ionization potentials. In order to
model this spectrum Netzer et al. (2003) invoked a photoionization
model which includes three ionized components, each split into two
kinematic components (at outflow velocities of 500 and 1000
km\,s$^{-1}$). The three components span a large range of ionization
and have total column density of about $4\times 10^{22}$ cm$^{-2}$. All
three components are thermally stable and seem to have the same gas
pressure.  Due to the detailed study of this spectrum it can serve as a
benchmark for the study of other high resolution X-ray spectra of AGNs.

In contrast to the many studies of the high-resolution X-ray
spectra of low-luminosity AGNs, there were only few such studies of
high-luminosity AGNs -- the quasars.  Studies by {\it ASCA} (e.g.,
George et al. 2000) show that the fraction of these with strong WAs
is below the equivalent fraction in low-luminosity AGNs. A plausible
explanation can be that the WA in high-luminosity AGNs is much more
ionized and thus escaped the detection by the low resolution of {\it
ASCA}, which is only sensitive to the strong bound-free absorption
by the O\,{\sc vii} and O\,{\sc viii} edges. To investigate this
hypothesis we are studying the X-ray high resolution grating spectra
of quasars from {\it Chandra} and {\it XMM-Newton} in search for
absorption lines. In this contribution we present some preliminary
results for three of the quasars we are studying.

\section{MR\,2251+178}

MR\,2251+178 ($z = 0.6398 \pm 0.00006, V \approx 14 $) is the first
quasar detected by X-ray observations. In Figure 1 we present the
high-resolution grating spectrum obtained by the RGS on May 2002. The
spectrum shows emission lines from N\,{\sc vi}, O\,{\sc vii}, O\,{\sc
viii}, Ne\,{\sc ix}, and Ne\,{\sc x}, as well as absorption lines
from O\,{\sc iii}, O\,{\sc iv}, O\,{\sc v}, O\,{\sc vi}, Ne\,{\sc ix}
and Ne\,{\sc x}.  Several other absorption lines which we suspect
their presence are marked in Figure 1.

\begin{figure}
\centerline{\includegraphics[width=13.0cm]{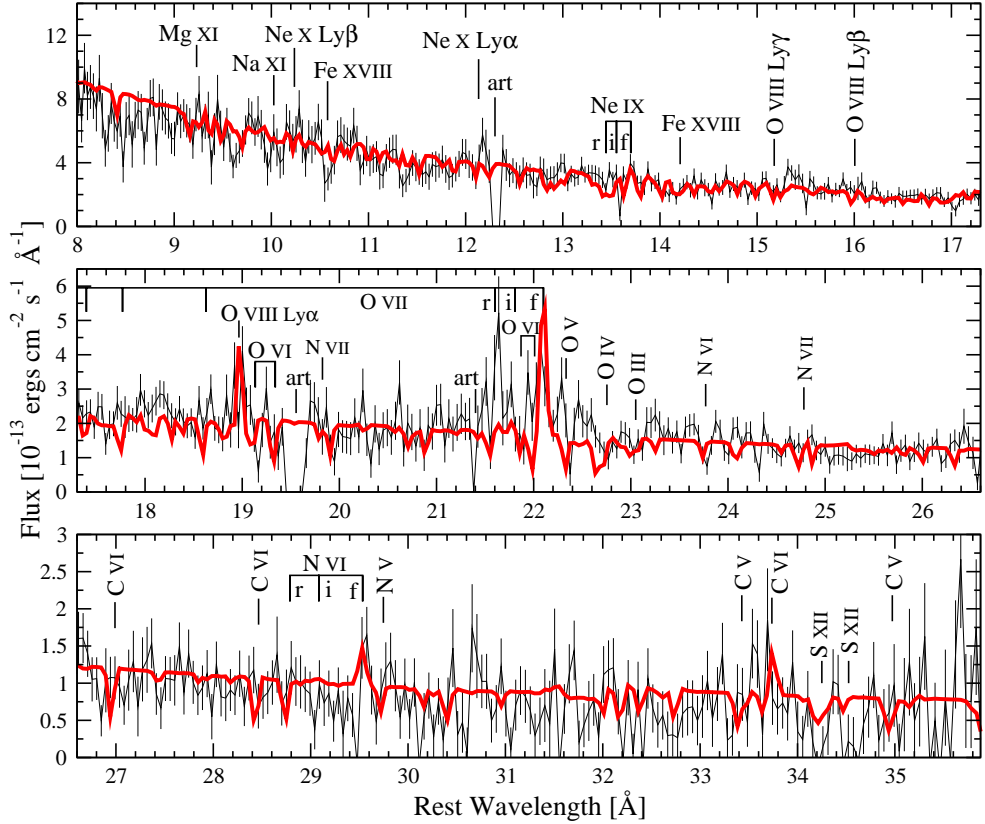}}
\caption{Combined RGS1 and RGS2 spectrum of MR\,2251+178 binned to
$\sim 0.04$
 \AA . The spectrum has been corrected for Galactic
absorption. The strongest emission lines are due to the O\,{\sc vii}
triplet and O\,{\sc viii} Ly$\alpha$. Other suggested emission and
absorption lines are marked.  Gaps in the spectrum due to chip gaps
are marked as `art'. The three absorbers model was convolved with
the RGS instrumental resolution and was also binned to 0.04~\AA .
\label{rgsspecmr} }
\end{figure}

In order to fit this high-resolution spectrum we first fitted the low
resolution EPIC-pn data obtained in the same observation.  The EPIC-pn
data clearly show a power law with a photon index of $\Gamma\approx
1.6$ at energies above 3 keV. Extrapolating this power law to lower
energies revels the presence of a WA around 0.8 keV, an additional
absorber below 0.5 keV, and some excess emission around 0.5 keV. Our
best fitted model for these data yields a WA with a column density,
$N_{\rm H}$, of $10^{21.51\pm0.03}$ cm$^{-2}$, ionization parameter,
$U_{\rm OX}$, of $10^{-1.78\pm0.05}$ and a line of sight covering
factor of 0.8. Assuming gas with the same properties produces the
emission, we find a global covering factor of 0.3. For the less
ionized absorber we find that it can be fitted with a neutral absorber
(in addition to the galactic one) with $N_{\rm H} \approx 10^{20.3}$
cm$^{-2}$ . This absorber can also be modeled as a combination of
low-ionization absorber with $U_{\rm OX} \approx 10^{-4}$ and $N_{\rm
H} \approx 10^{20.3}$ cm$^{-2}$\ and a neutral absorber of $N_{\rm H}
\approx 10^{20.06}$ cm$^{-2}$ . Both cases give equally good fits.

Modeling of the RGS spectrum was done in two steps. First
we experimented with a two-components absorber similar to the
one fitted to the EPIC-pn data described above, which involves a
highly-ionized absorber and a second absorber of much lower-ionization.
We experimented with a range of parameters around the values
found for the EPIC-pn. The parameters we found to fit best for
the two absorbers in the RGS spectrum are: log($U_{\rm OX})=-1.68$
and $N_{\rm H}=10^{21.8}$ cm$^{-2}$\ for the high-ionization WA component
and log($U_{\rm OX})=-4.0$ and $N_{\rm H}=10^{20.3}$ cm$^{-2}$\ for the
low-ionization component.  The second step includes a three-components
absorber. The main motivation for this is the fact that the RGS data
around 16--17~\AA\ clearly falls below the two-components model. The
excess absorption is probably caused by the unresolved transition array
(UTA) of iron M-shell lines which has been observed in several other
AGNs (see, e.g., Netzer et al. 2003 for the case of NGC\,3783). Our
photoionization code includes all these lines but the two-components WA
produces too shallow a feature at too short a wavelength. We find that
an additional shell with a column density of $10^{21.3}$ cm$^{-2}$\
and log$(U_{\rm OX})=-2.6$ can significantly improve the fit. This
component produces a noticeable UTA feature and contributes also to
the observed O\,{\sc vii} emission. This requires lowering the emission
from the high-ionization component by about 20\% to produce an adequate
fit to all emission lines.  Adding this component force us to increase
the ionization parameter of the highly ionized gas (the one with column
density of $10^{21.8}$ cm$^{-2}$ ) to log$(U_{\rm OX})=-1.4$. We note
that the mean $U_{\rm OX}$ of these two WA components is the same as
the one found earlier in the two-components model. The three-components
model is compared with the RGS data in Figure 1.

\section{PG\,1211+143}

PG\,1211+143 ($z = 0.0809 \pm 0.0005, V \approx 14.5 $) is a
radio-quiet narrow emission line quasar. It was observed with {\it
XMM-Newton} for about 60 ks on June 2001. Using the EPIC-pn, EPIC-MOS,
and RGS data Pounds et al. (2003) identified absorption lines from
H-like and He-like ions of Fe, S, Mg, Ne, O, N, and C. They claimed the
observed line energies indicate an ionized outflow velocity of $\sim
24000$ km\,s$^{-1}$. We obtained the data from the {\it XMM-Newton}
archive and present in Figure 2 the RGS data.  Our preliminary
re-analysis of the data shows that an ionized absorber with outflow
velocity of about 3000 km\,s$^{-1}$ provides an alternative explanation
to the data. We are able to identify at this velocity a series of
absorption lines from H-like and He-like ions of Mg, Ne, and O, as
well as L-shell lines from iron ions.  We also include the emission
seen from the O\,{\sc vii} and Ne\,{\sc ix} forbidden lines, and the
Ly$\alpha$ emission lines of O\,{\sc viii} and Ne\,{\sc x}. These
emission lines together with the many L-shell iron lines were not
taken into account in the model by Pounds et al. (2003).

\begin{figure}
\centerline{\includegraphics[width=13.0cm]{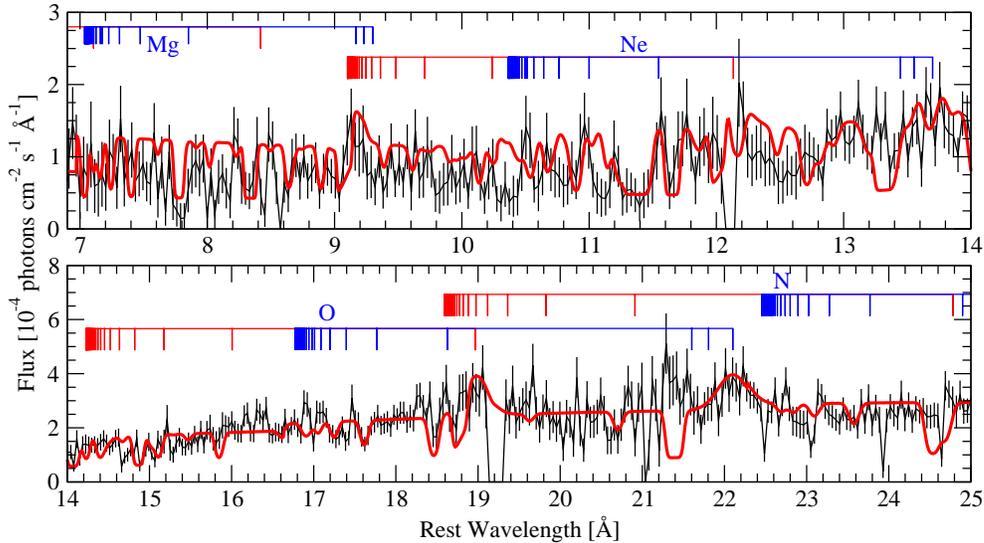}}
\vglue -1.4cm
\caption{Combined RGS1 and RGS2 spectrum of PG\,1211+143 binned
to $\sim 0.04$ \AA . The spectrum has been corrected for Galactic
absorption. The model shown includes absorption lines blueshifted
by 3000 km\,s$^{-1}$ and few emission lines. Theoretical rest-frame
wavelengths of H-like and He-like ions of N, O, Ne, and Mg are marked
above the spectrum.
\label{rgsspecpg} }
\end{figure}

\section{4C\,74.26}

4C\,74.26 ($z = 0.104, V \approx 14.8 $) is a radio loud quasar
with relatively strong X-ray flux compared with other quasars. It
was observed with {\it Chandra}/HETGS on December 2003 for 70 ks. A
preliminary analysis of the data revels emission and absorption
lines from H-like and He-like ions of Mg, Al, and Si, as well as
lower ionization Si absorption lines. These data, as well as the ones
presented in the previous sections, indicate that highly-ionized WAs
are present in quasars and can be detected with the power of high
resolution X-ray spectroscopy.

\begin{acknowledgments}

I am grateful to my collaborators in the studies presented here,
Hagai Netzer, Ehud Behar, Doron Chelouche, Ian, M. George, Kirpal
Nandra, and T.J. Turner.  I acknowledge a financial support by the Israel
Science Foundation grant no. 545/00.

\end{acknowledgments}


\begin{thebibliography}{}

\bibitem[Behar et al.(2003)]{2003ApJ...598..232B} Behar, E., et al.
2003, ApJ, 598, 232 

\bibitem[Collinge et al.(2001)]{2001ApJ...557....2C} Collinge, M.~J., et al.
\ 2001, ApJ, 557, 2 

\bibitem[George et al.(1998)]{1998ApJS..114...73G} George, I.~M., et al.
1998, ApJS, 114, 73 

\bibitem[George et al.(2000)]{2000ApJ...531...52G} George, et al.
2000, ApJ, 531, 52 

\bibitem[Kaspi et al.(2002)]{2002ApJ...574..643K} Kaspi, S., et al.\ 2002, 
ApJ, 574, 643 

\bibitem[Netzer et al.(2003)]{2003ApJ...599..933N} Netzer, H., et al.\ 
2003, ApJ, 599, 933 

\bibitem[Pounds et al.(2003)]{2003MNRAS.345..705P} Pounds, K.~A., et al.
2003, MNRAS, 345, 705 

\bibitem[Steenbrugge, Kaastra, de Vries, \& 
Edelson(2003)]{2003A&A...402..477S} Steenbrugge, K.~C.,
Kaastra, J.~S., de Vries, C.~P., \& Edelson, R.\ 
2003, A\&A, 402, 477 

\end{thebibliography}
\end{document}